\documentclass[
reprint,longbibliography,
superscriptaddress,
 amsmath,amssymb,
prl,
]{revtex4-2}

\newcommand\beq{\begin{equation}}
\newcommand\eeq{\end{equation}}

\usepackage{graphicx}
\usepackage{amsmath}
\usepackage{amsfonts}
\usepackage{color}
\usepackage[normalem]{ulem}
\usepackage[colorlinks, linkcolor= blue, citecolor = blue, urlcolor=blue]{hyperref}

\begin{document}


\title{Analytic Inverse Design of Temporal Metamaterials\\via Space-Time Duality}

\author{Giuseppe Castaldi}
\affiliation{Fields \& Waves Lab, Department of Engineering, University of Sannio, I-82100 Benevento, Italy}
\author{Marino Coppolaro}
\affiliation{Fields \& Waves Lab, Department of Engineering, University of Sannio, I-82100 Benevento, Italy}
\author{Massimo Moccia}
\affiliation{Fields \& Waves Lab, Department of Engineering, University of Sannio, I-82100 Benevento, Italy}
\author{Carlo Rizza}
\affiliation{Department of Physical and Chemical Sciences, University of L’Aquila, I-67100 L’Aquila, Italy}
\author{Nader Engheta}
\affiliation{Department of Electrical and Systems Engineering, University of Pennsylvania, Philadelphia, Pennsylvania 19104, USA}
\author{Vincenzo Galdi}
\email{vgaldi@unisannio.it}
\affiliation{Fields \& Waves Lab, Department of Engineering, University of Sannio, I-82100 Benevento, Italy}

\date{\today}


\begin{abstract}
Temporal metamaterials, created by modulating the refractive index in time, offer powerful means of controlling wave propagation but still lack a systematic design methodology. Here, we develop an analytic inverse-design framework rooted in space-time duality and the established theory of one-dimensional spatial inverse scattering. By prescribing reflection (backward-wave) and transmission (forward-wave) responses in rational-function form, we obtain closed-form refractive-index modulations that are guaranteed to be physically admissible. This approach avoids iterative optimization and provides direct analytic control of the modulation. We illustrate the method with syntheses of mathematical operators, such as derivatives and integrals, as well as Chebyshev- and Butterworth-type filters, and validate the results through finite-difference time-domain simulations. Our findings establish a general route to temporal media with tailored functional and spectral responses, enabling applications in wave-based information processing, programmable filtering, and amplification schemes inspired by photonic time crystals.
\end{abstract}

\maketitle

Time-varying electromagnetic (EM) media have been a subject of interest since the 1950s, with early works investigating wave propagation in temporally modulated dielectric and plasma systems \cite{Morgenthaler:1958vm,Oliner:1961wp,Felsen:1970wp,Fante:1971to}. These foundational works established key principles of nonstationary interactions between EM waves and dynamic material properties. In recent years, advancements in metamaterials science and engineering have reignited interest in this field, enabling precise and rapid modulation of constitutive properties in both space and time, thereby adding a new dimension to the design space \cite{Engheta:2021mw,Engheta:2023fd}. This resurgence has led to a broad range of novel applications, including nonreciprocal wave propagation, time-refraction,   frequency conversion,  and wave manipulation beyond 
the constraints of conventional  EM systems (see \cite{Galiffi:2022po} for a recent review).
 
Most design strategies for temporal metamaterials leverage space-time duality and draw inspiration from spatial-domain synthesis concepts. For example, stepped-transmission-line analogies have been used to realize temporal multilayers \cite{Ramaccia:2021tm,Castaldi:2021es,Castaldi:2022he,Silbiger:2025id}, while tapering techniques have enabled soft-switching temporal profiles described by Riccati-type equations \cite{Galiffi:2022tp,Silbiger:2023of,Silbiger:2024id}. In a complementary and more task-specific direction, short-pulsed temporal modulations have been analytically engineered to implement selected elementary operations (e.g., derivatives), rather than providing a general synthesis framework \cite{Rizza:2022sp}. These operations can then serve as building blocks for more complex responses through temporal-stacking mechanisms inspired by the cascading principles of metasurfaces \cite{Castaldi:2023ma}.

However, none of these approaches fully harnesses the potential of space-time duality, particularly the rigorous and systematic design methodologies established for the inverse design of one-dimensional (1D) spatially varying media. In this work, we aim to close this gap by adapting and extending these powerful techniques to the temporal domain.
 
 %
 \begin{figure}[t]
 	\centering
 	\includegraphics[width=\linewidth]{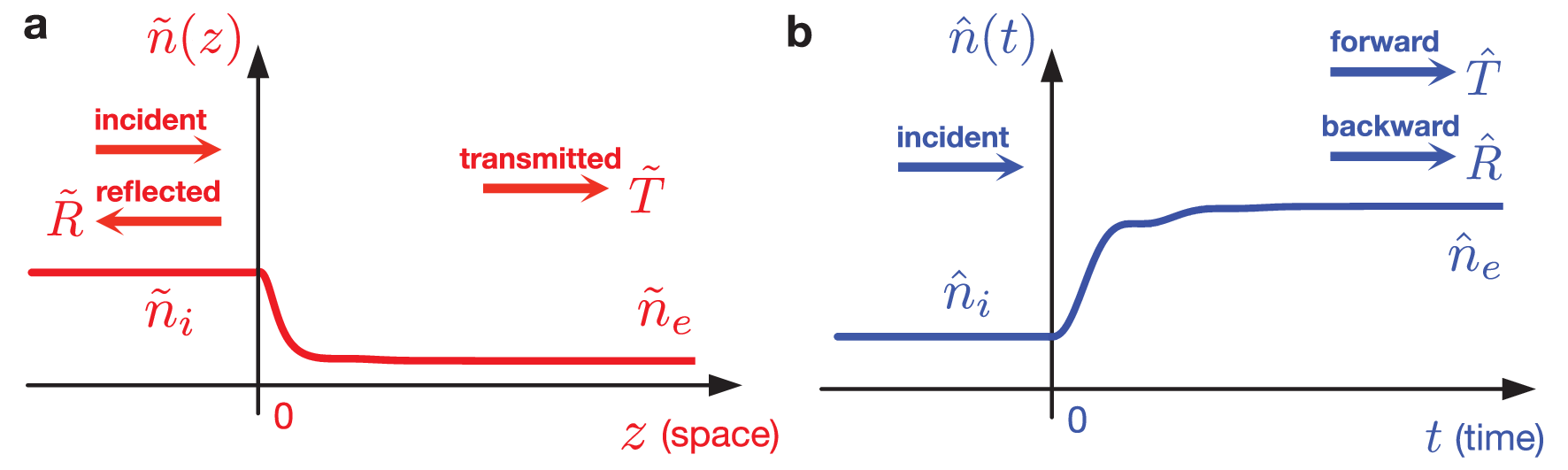}
 	\caption{Conceptual illustration of the space-time duality. (a), (b) Wave interactions with a spatially inhomoheneous and  time-varying medium, respectively. Relevant quantities and details are described in the text.}
 	\label{Figure1}
 \end{figure}
 
The theoretical foundations of inverse scattering problems for 1D spatial media, aimed at reconstructing the spatial inhomogeneity profile from its reflection response, were established in the 1950s through the pioneering work of Gel’fand, Levitan, and Marchenko \cite{Gelfand:1955at,Marchenko:1955ro} in quantum mechanics, formulated in terms of integral equations. 
Closely related spectral formulations of inverse scattering were later introduced by Zakharov \& Shabat in the context of integrable nonlinear wave equations \cite{Zakharov:1972et}. 
These methodologies were later adapted to EM scenarios by Kay \cite{Kay:1960ti} and Moses \& de Ridder \cite{Moses:1963po}, 
leading to extensive numerical and analytical advancements through the late 1980s (see \cite{Jaggard:1987te} for a comprehensive review). A major breakthrough in this domain came with an alternative approach by Balanis \cite{Balanis:1982is}, which mitigated some complexities and limitations of the Gel’fand-Levitan-Marchenko framework. This method enhanced numerical stability and enabled systematic analytical reconstruction in specific cases, such as when the reflection coefficient is a rational function of frequency \cite{Knockaert:2000ar}.

In the following, we demonstrate how this approach can be systematically extended to the inverse design of temporal metamaterials, enabling the analytical retrieval of time-varying refractive index profiles that produce desired reflection (backward-wave) or transmission (forward-wave) responses, parameterized as rational functions of the wavenumber.
 
 Figure \ref{Figure1} conceptually illustrates the space-time duality underlying our approach. Specifically, Fig. \ref{Figure1}a depicts the spatial scenario, where a refractive index distribution $\tilde{n}(z)$ varies inhomogeneously along a 1D medium. The refractive index remains constant at $\tilde{n}_i$ in the half-space $z < 0$, and asymptotically approaches $\tilde{n}_e$ as $z \to \infty$. Here and henceforth, the tilde ($\tilde{~}$) denotes spatial-domain quantities.
 Assuming a time-harmonic plane-wave incidence  from $z < 0$ with time-harmonic dependence $\exp(-i\omega t)$ (where $\omega$ denotes the conserved angular frequency), we define the reflection and transmission coefficients, $\tilde{R}(\omega)$ and $\tilde{T}(\omega)$, respectively (see \cite{SM} for details). In what follows, we neglect dispersion and losses, assuming operation far from resonances. These assumptions, along with causality, imply that the reflection and transmission coefficients must satisfy the constraints  $\tilde{F}(-\omega) = \tilde{F}^*(\omega)$ (with $\tilde{F}=\tilde{R},\tilde{T}$) and  $|\tilde{R}(\omega)| \leq 1$, and cannot have poles in the upper half-plane $\mbox{Im}(\omega) > 0$.
 Finally, power conservation establishes the relationship
 \beq
 \left|\tilde{T}\left(\omega\right)\right|^2=
 \frac{\tilde{n}_i}{\tilde{n}_e}\left[1-\left|\tilde{R}\left(\omega\right)\right|^2\right].
 \label{eq:PCS}
 \eeq 
  Figure \ref{Figure1}b illustrates the temporal scenario, where a refractive index $\hat{n}(t)$ varies over time, starting  from the constant value $\hat{n}_i$ for $t < 0$, and asymptotically approaching $\hat{n}_e$ as $t \to \infty$. Here and henceforth, the caret (\^{}) denotes temporal-domain quantities.
  In this context, we consider a time-harmonic plane wave of the form $\exp(i k z)$, where $k$ represents the conserved wavenumber. This allows us to define the temporal reflection (backward-wave) and transmission (forward-wave) coefficients, $\hat{R}(k)$ and $\hat{T}(k)$, respectively.
  Similar to the spatial case, the Hermitian symmetry condition $\hat{F}(-k) = \hat{F}^*(k)$ holds (with $\hat{F}=\hat{R},\hat{T}$), and physical feasibility requires that $\hat{R}(k)$ and $\hat{T}(k)$ have no poles in the lower half-plane $\mbox{Im}(k) < 0$.
  However, unlike the spatial scenario, power conservation does not generally apply in the temporal case, as the system can exchange energy with the modulation. The relationship between the reflection and transmission coefficients is given by:
 \beq
 \left|\hat{T}(k)\right|^2=\frac{{\hat{n}}_e}{{\hat{n}}_i}+\left|\hat{R}\left(k\right)\right|^2.
 \label{eq:PCT}
 \eeq
  It can be shown (see \cite{SM} for details) that the spatial and temporal scenarios become {\em mathematically equivalent} under the following duality transformations:  
  \beq
  z\rightarrow ct,\quad kc\rightarrow-\omega,  \quad {\hat n}\left(t\right)= \frac{1}{{\tilde n}\left(z\rightarrow c t\right)},
  \label{eq:transf}
  \eeq
  with the additional correspondence that the electric induction $D$ in the spatial formulation is mapped onto the magnetic induction $B$ in the temporal formulation.
  Here and henceforth, $c$ denotes the wavespeed in vacuum.
  These transformations establish that both cases are governed by {\em identical} wave-transport operators, ensuring a direct correspondence between their scattering properties. By applying the space-time duality transformation and systematically comparing the temporal and spatial scattering problems, which are characterized by different boundary conditions (see Fig. \ref{Figure1}), we find that the reflection and transmission responses in the two domains are related by the following expression (see \cite{SM} for details):
  \beq
  \hat{R}\left(-\frac{\omega}{c}\right)=-\frac{{\hat{n}}_e \tilde{R}\left(\omega\right)}{{\hat{n}}_i\tilde{T}\left(\omega\right)}.
  \label{eq:Rk1}
  \eeq
  This fundamental relationship serves as the cornerstone of our study, enabling the systematic extension of design principles from one domain to the other.
  
  In particular, we assume a prescribed temporal reflection coefficient expressed as a rational function of the wavenumber,  
 \beq
 \hat{R}\left(k\right)=\frac{P_N\left(k\right)}{Q_M\left(k\right)},
 \label{eq:RIk}
 \eeq
  where $P_N(k)$ and $Q_M(k)$ are polynomials of degree $N$ and $M$, respectively, and the zeros of $Q_M(k)$ have positive imaginary part.
  To ensure that in the static limit ($k\rightarrow 0$, i.e., step variation) the system consistently reduces to a canonical temporal boundary \cite{Xiao:2014ra}, we further enforce the condition
  \beq
  \hat{R}\left(0\right)=\frac{{\hat n}_i-{\hat n}_e}{2{\hat n}_i}.
  \label{eq:R0}
  \eeq
  
  Accordingly, the inverse design of the time-varying refractive index can be formulated as an equivalent spatial-domain problem, where the corresponding spatial reflection coefficient  is given by (see \cite{SM} for details):  
 \beq
 \tilde{R}\left(\omega\right)=\frac{ P_N\left(\displaystyle{-\frac{\omega}{c}}\right)}{\displaystyle{\prod_{m=1}^{M}}\left(\omega-\omega_{rm}\right)},
 \label{eq:Rw1}
 \eeq
  where $\omega_{rm}$ are the zeros of $|\hat{R}(-\omega/c)|^2 + \hat{n}_e / \hat{n}_i$ that have a negative imaginary part, and we enforce  the self-consistency condition
  \beq
  {\tilde R}\left(0\right)=\frac{{\tilde n}_i-{\tilde n}_e}{{\tilde n}_i+{\tilde n}_e}=\frac{{\hat n}_e-{\hat n}_i}{{\hat n}_e+{\hat n}_i}.
  \label{eq:R0}
  \eeq
  
  Similarly, if we instead prescribe the transmission coefficient as a rational function of the wavenumber,  
  \beq
  \hat{T}(k) = \frac{U_N(k)}{V_M(k)},\quad  \hat{T}(0)=\frac{{\hat n}_i+{\hat n}_e}{2{\hat n}_i},
  \label{eq:Tk}
  \eeq  
 where $U_N(k)$ and $V_M(k)$ are polynomials of degree $N$ and $M$, respectively, the zeros of $V_M(k)$ have positive imaginary part, and a self-consistency condition in the static limit \cite{Xiao:2014ra} has been enforced, the corresponding spatial reflection coefficient is given by (see \cite{SM} for details):  
  \beq
  \tilde{R}(\omega) = \frac{\displaystyle{\prod_{m=1}^{M}} (\omega - \omega_{tm})}{U_N\left(\displaystyle{-\frac{\omega}{c}}\right)}.
  \label{eq:Rw2}
  \eeq  
  Here, the zeros of $U_N(-\omega/c)$ are selected with negative imaginary part,
  $\omega_{tm}$ denote the zeros of $|\hat{T}(-\omega/c)|^2 - \hat{n}_e / \hat{n}_i$ that also lie in the
  lower half of the complex $\omega$-plane, and the self-consistency relationship in (\ref{eq:R0}) is enforced.

 In summary, the synthesis of temporal metamaterials with prescribed reflection or transmission responses, expressed in rational-function form, can be systematically mapped onto equivalent spatial inverse-scattering problems. These spatial formulations admit analytical closed-form solutions \cite{Knockaert:2000ar} (see also \cite{SM} for details), thereby providing a rigorous and versatile framework for inverse design. Once the spatial refractive-index profile $\tilde n(z)$ is obtained, the corresponding temporal profile ${\hat n}(t)$ follows directly from the third of the space-time duality transformations in (\ref{eq:transf}).

In what follows, we present and analyze representative synthesis examples, with numerical validation carried out using a finite-difference time-domain (FDTD) solver \cite{Oskooi:2010ma}. In these examples, the syntheses are formulated by prescribing the reflection response, while the transmission response is not independently designed but follows from the coupled scattering relations in Eq. (\ref{eq:PCT}).
For conciseness, the full expressions of the synthesized operators and excitation waveforms, along with implementation details of the syntheses and numerical simulations, are provided in \cite{SM}.
 
 %
 \begin{figure}[t]
 	\centering
 	\includegraphics[width=\linewidth]{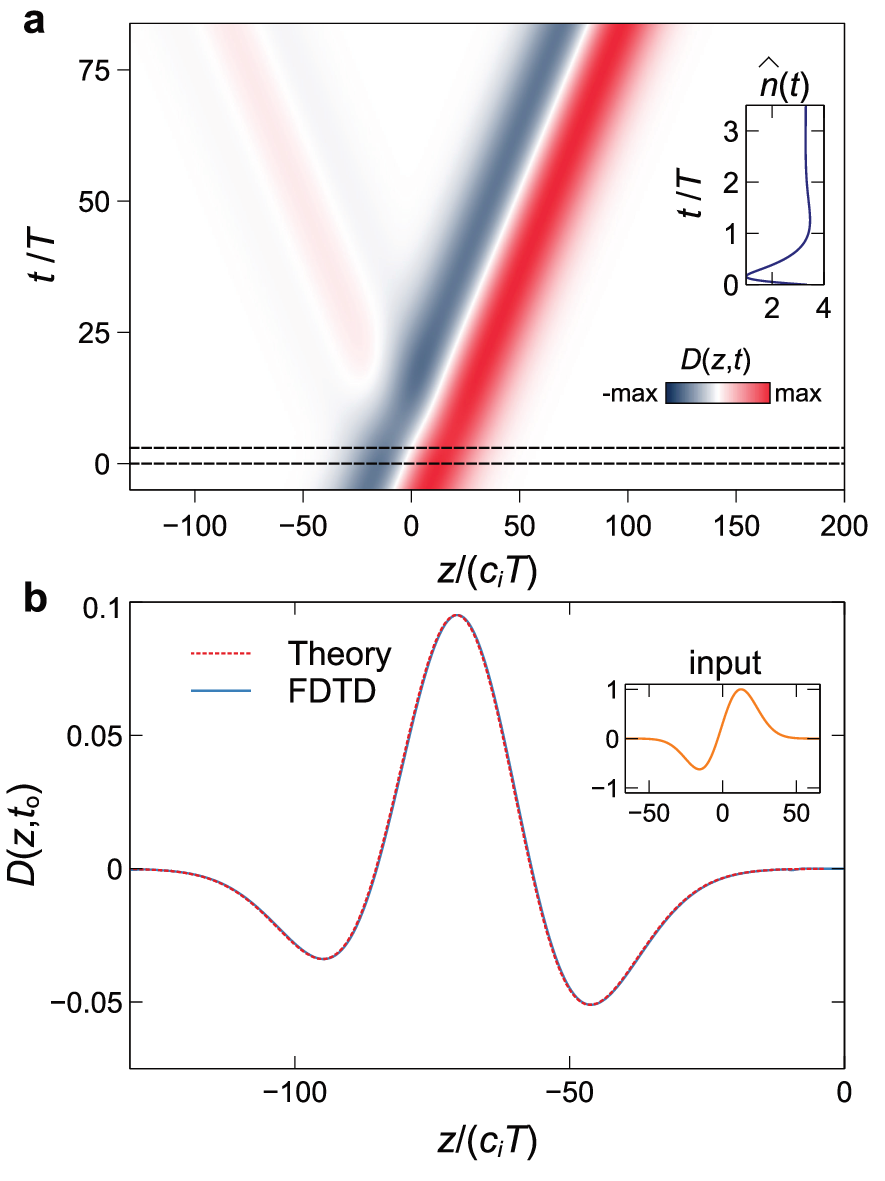}
 	\caption{(a) FDTD-computed space-time map of  normalized electric displacement field for a temporal metamaterial synthesized to implement a first-derivative operation in the reflection (backward) response, assuming $\hat{n}_i =\hat{n}_e= 3.3$. The modulation interval is marked by horizontal black dashed lines, and the inset shows the corresponding time-varying refractive-index profile $\hat{n}(t)$. 
 		(b) Spatial field distribution at $t_o = 70T$, comparing FDTD simulation results with the theoretical prediction. The inset displays the modulated-Gaussian wavepacket used as the incident excitation.}
 	\label{Figure2}
 \end{figure}
 
 As a first example, Fig. \ref{Figure2} reports the synthesis of a band-limited first-derivative operator in the reflection (backward-wave) response. The numerically computed space-time distribution of the electric displacement field is shown in Fig. \ref{Figure2}a, with the corresponding synthesized refractive-index profile $\hat{n}(t)$ displayed in the inset. As observed, the refractive-index modulation is effectively confined to a finite temporal interval (with $\hat{n}_i = \hat{n}_e = 3.3$), and the synthesis is constrained so that  $\hat{n}(t) \geq 1$ (see \cite{SM} for details). 
 Figure \ref{Figure2}b compares the numerically computed response (evaluated at $t_o= 70T$) with the exact theoretical prediction, for a modulated-Gaussian input wavepacket shown in the inset. Note that the input signal is chosen to be slightly asymmetric, in order to avoid symmetry-induced ambiguities. Here and henceforth, $T$ denotes a reference timescale used for normalization, and $c_i=c/{\hat n}_i$ is the speed of light in the initial medium. As can be seen, the agreement between simulation and theory is excellent.
 
Additional synthesis examples of band-limited mathematical operations, including the second derivative and the integral, are reported in \cite{SM}. 

These results can be regarded as an extension of the approach in Ref. \cite{Rizza:2022sp}, providing a more systematic and versatile synthesis framework, and operating without constraints on the modulation duration.

%
\begin{figure}[t]
	\centering
	\includegraphics[width=\linewidth]{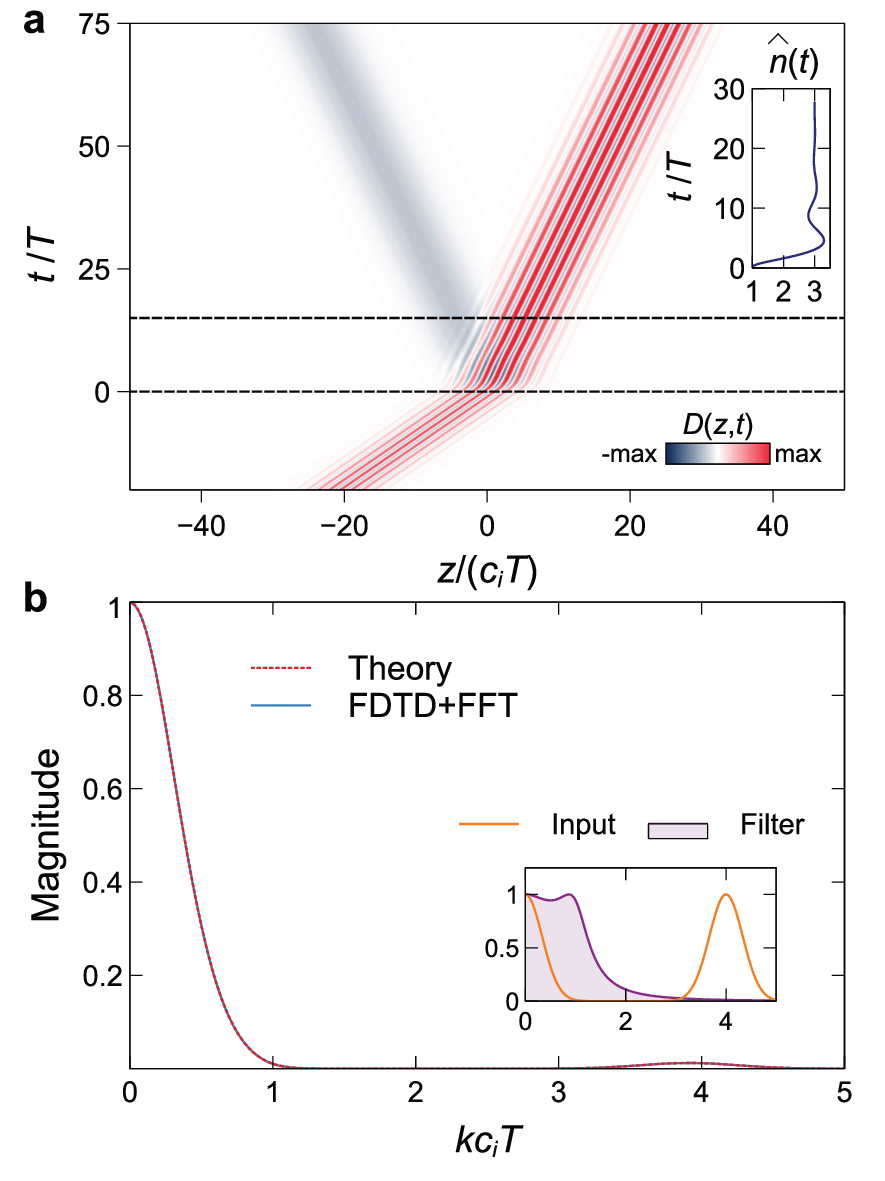}
	\caption{(a) FDTD-computed space-time map of  normalized electric displacement field for a temporal metamaterial synthesized to implement a Chebyshev-type low-pass filter in the reflection (backward-wave) response, assuming ${\hat n}_i=1$ and ${\hat n}_e=3$. The modulation interval is marked by horizontal black dashed lines, and the inset shows the corresponding time-varying refractive-index profile $\hat{n}(t)$. 
		(b) Wavenumber-spectrum (magnitude) of normalized electric displacement field at $t = 70T$, comparing FDTD simulation results (transformed via FFT) with the theoretical prediction. The inset displays the incident spectrum and the filter response.}
	\label{Figure3}
\end{figure}

As a second representative example, Fig. \ref{Figure3} illustrates the synthesis of a low-pass Chebyshev-type (i.e., equiripple) filter \cite{Williams:2006ef} in the reflection (backward-wave) response. Figure \ref{Figure3}a shows the numerically computed space–time distribution, with the corresponding synthesized refractive-index profile $\hat{n}(t)$ displayed in the inset. As in the previous case, the refractive-index modulation is confined to a finite temporal interval and remains above unity; however, here the initial and final values differ ($\hat{n}_i = 1, \hat{n}_e = 3)$. This feature directly follows from (\ref{eq:R0}), ensuring a nonzero static response consistent with the low-pass character of the synthesized filter.
For this example, a spectral comparison is more informative than a time-domain one. Accordingly, Fig. \ref{Figure3}b compares the exact spectral wavenumber response (at a fixed time) with the numerical counterpart obtained via fast Fourier transform (FFT) of the FDTD-computed signal. The inset shows both the filter response and the spectrum of the input, which consists of two modulated-Gaussian wavepackets (one within the filter passband and the other outside it). Also in this case, the agreement between theory and simulation is excellent.

A related synthesis, implementing a Chebyshev-type passband filter, is reported in \cite{SM}.

%
\begin{figure}[t]
	\centering
	\includegraphics[width=\linewidth]{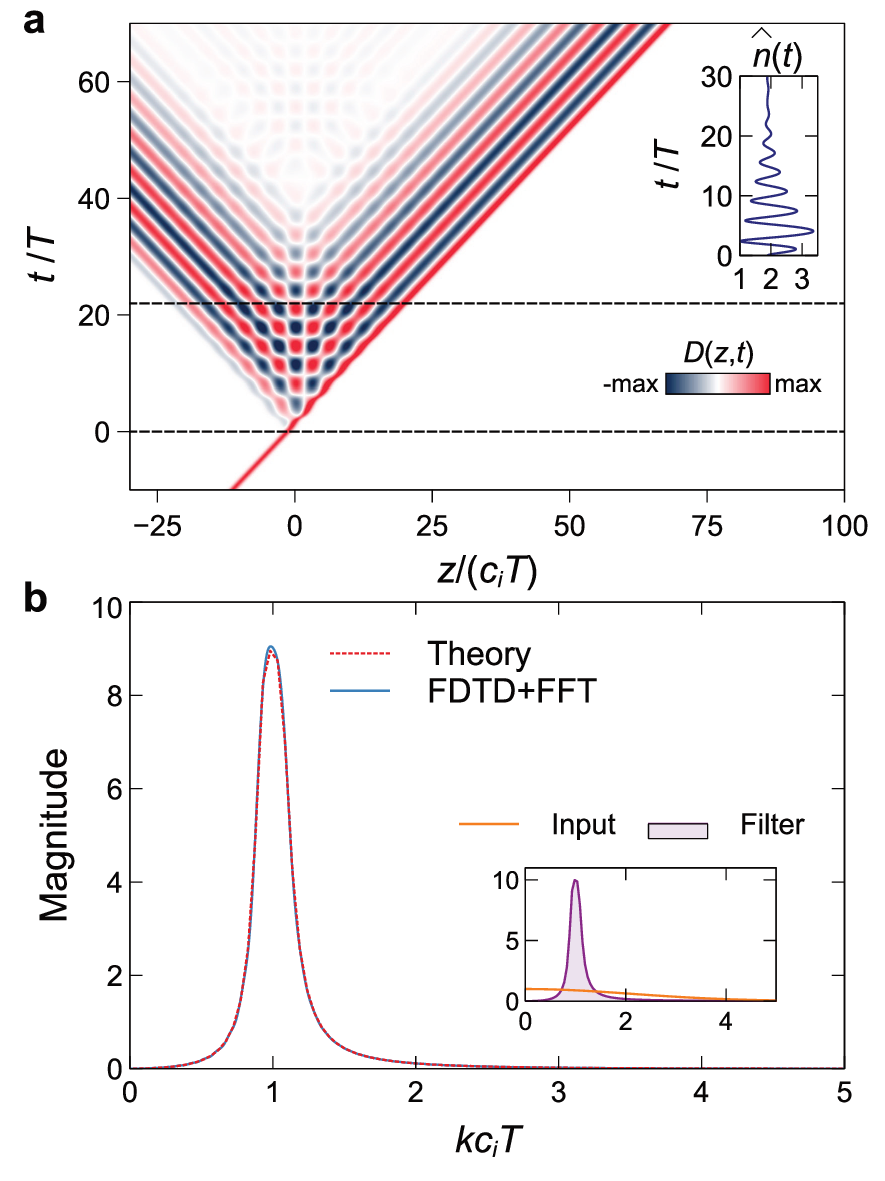}
	\caption{(a) FDTD-computed space-time map of normalized electric displacement field for a temporal metamaterial synthesized to implement a Butterworth-type narrow-band amplifying filter in the reflection (backward-wave) response, assuming ${\hat n}_i={\hat n}_e=2$. The modulation interval is marked by horizontal black dashed lines, and the inset shows the corresponding time-varying refractive-index profile $\hat{n}(t)$. 
		(b) Wavenumber-spectrum (magnitude) of normalized electric displacement field at $t = 70T$, comparing FDTD simulation results (transformed via FFT) with the theoretical prediction. The inset displays the incident spectrum and the filter response.}
	\label{Figure4}
\end{figure}

As a final example, we consider the implementation of a narrowband Butterworth-type (i.e., maximally flat) amplifying filter \cite{Williams:2006ef} in the reflection (backward-wave) response, with results summarized in Fig. \ref{Figure4}. The space–time map in Fig. \ref{Figure4}a shows that both backward- and forward-wave spectra exhibit a pronounced, narrowband enhancement. The corresponding refractive-index distribution, displayed in the inset, features a quasi-periodic, damped modulation that remains  above unity and returns to the same value at the beginning and end of the modulation interval ($\hat{n}_i = \hat{n}_e = 2$). This behavior is consistent with the amplification mechanisms inherent in {\em photonic time crystals} \cite{Martinez:2016tp,Lustig:2023pt,Asgari:2024ta}, where periodic temporal modulations give rise to momentum bandgaps. In contrast, our synthesis framework enables precise and direct control over the amplification level, bandwidth, and roll-off characteristics.
Figure \ref{Figure4}b presents the corresponding spectral response, once again demonstrating excellent agreement between theoretical predictions and FDTD simulations.

It is worth noting that, in Figs. \ref{Figure2} and \ref{Figure3}, the engineered reflection responses remain moderate and do not involve amplification; consequently, only a limited redistribution of spectral content occurs, and the transmission coefficient stays close to unity over the relevant bandwidth, leaving the transmitted waveform only weakly modified. By contrast, in Fig. \ref{Figure4}, the narrowband enhancement engineered in reflection introduces a stronger spectral selectivity which, through the intrinsic coupling between reflection and transmission, results in a more pronounced reshaping and temporal broadening of the transmitted signal.
The reader is also referred to \cite{SM} for a complementary example in which the synthesis directly targets the transmission (forward-wave) response.

In conclusion, we have introduced a rigorous and systematic framework for the analytic inverse design of temporal metamaterials, grounded in space-time duality and the theory of 1D spatial inverse scattering. By prescribing reflection or transmission responses in rational-function form, we demonstrated the direct retrieval of physically admissible refractive-index modulations. Representative syntheses, including mathematical operators (derivatives, integrals) and advanced filter functions (Chebyshev, Butterworth), were validated numerically with excellent agreement between theory and simulation.

Beyond establishing fundamental design principles, our approach provides a versatile route to time-varying media with tailored functionalities, avoiding iterative optimization and enabling analytic control over key parameters. In contrast to other analytic-based methods \cite{Galiffi:2022tp,Silbiger:2023of}, it does not rely on small-reflection assumptions, thereby allowing rigorous treatments of more general scattering scenarios. These capabilities open opportunities in temporal analog computing, dynamic filtering, amplification schemes inspired by photonic time crystals, and hybrid space-time metamaterials with unprecedented control over wave dynamics.

Future directions may include extending the framework to dispersive and lossy media, as well as exploring temporally varying profiles combined with spatial structuring, thereby broadening the scope of inverse-designed temporal metamaterials for photonics, communications, and wave-based information processing. In this context, it is worth noting that inverse-scattering formulations for dispersive media have been developed in the spatial domain (e.g., for stratified plasmas) \cite{Balanis:1972tp}, suggesting a possible route to extend the present approach to temporally dispersive systems. It would also be interesting to further investigate the implications of the proposed space-time duality framework in light of recent studies showing that phenomena typically associated with photonic time crystals can arise in spatially periodic systems subjected to temporal perturbations \cite{Li:2025tt}.

~\\
\begin{acknowledgments}
{\em Acknowledgments} --
	This work was supported in part by the European Union-Next Generation EU under the Italian National Recovery and Resilience Plan (NRRP), Mission 4, Component 2, Investment 1.3, CUP E63C22002040007, partnership on ``Telecommunications of the Future'' under Grant PE00000001 - program ``RESTART'', and in part from the University of Sannio via the FRA 2024 program.
	N. E. acknowledges partial support from the Simons Foundation/Collaboration on Symmetry-Driven Extreme Wave Phenomena (grant SFI-MPSEWP-00008530-04).
\end{acknowledgments}


%

\end{document}